\documentclass[10pt,conference]{IEEEtran}
\IEEEoverridecommandlockouts
\usepackage{cite}
\usepackage{amsmath,amssymb,amsfonts}
\usepackage{algorithmic}
\usepackage{graphicx}
\usepackage{textcomp}
\usepackage{xcolor}
\usepackage{physics}
\usepackage{array}
\usepackage{comment}
\usepackage[ruled,vlined]{algorithm2e}
    \newcolumntype{P}[1]{>{\centering\arraybackslash}p{#1}}
    \newcolumntype{M}[1]{>{\centering\arraybackslash}m{#1}}
\def\BibTeX{{\rm B\kern-.05em{\sc i\kern-.025em b}\kern-.08em
    T\kern-.1667em\lower.7ex\hbox{E}\kern-.125emX}}
\begin{document}

\title{Secured Quantum Identity Authentication Protocol for Quantum Networks \\\

\vspace{-12mm}

\author{Mohamed Shaban$^{\S,\dagger}$ and Muhammad Ismail$^\S$\\

{$^\S$Department of Computer Science, Tennessee Technological University, Cookeville, TN, USA}\\
{$^\dagger$Department of Mathematics, Faculty of Education, Alexandria University, Egypt} \\

Emails: \{mmibrahims42, mismail\}@tntech.edu

\thanks{This work was supported by NSF Award \# 2210251.}
}
}

\maketitle

\begin{abstract}
Quantum Internet signifies a remarkable advancement in communication technology, harnessing the principles of quantum entanglement and superposition to facilitate unparalleled levels of security and efficient computations. Quantum communication can be achieved through the utilization of quantum entanglement. Through the exchange of entangled pairs between two entities, quantum communication becomes feasible, enabled by the process of quantum teleportation.
 Given the lossy nature of the channels and the exponential decoherence of the transmitted photons, a set of intermediate nodes can serve as quantum repeaters to perform entanglement swapping and directly entangle two distant nodes. Such quantum repeaters may be malicious and by setting up malicious entanglements, intermediate nodes can jeopardize the confidentiality of the quantum information exchanged between the two communication nodes. Hence, this paper proposes a quantum identity authentication protocol that protects quantum networks from malicious entanglements. Unlike the existing protocols, the proposed quantum authentication protocol does not require periodic refreshments of the shared secret keys. Simulation results demonstrate that the proposed protocol can detect malicious entanglements with a $100\%$ probability after an average of $4$ authentication rounds.     
\end{abstract}


\section{Introduction}      
\IEEEPARstart{Q}{uantum} Internet represents the next evolutionary leap in communication technology, utilizing the principles of quantum entanglement and superposition to enable unprecedented levels of security and efficient computations. Quantum communications can be enabled via quantum entanglement. By sharing an entangled pair between two parties, quantum communications can be carried out over the classical Internet via quantum teleportation. This eliminates the requirement of having direct quantum channels between every two nodes.
 To initiate quantum communication, an entangled pair of photons can be generated and transmitted over a quantum channel to communication parties. 
 However, these channels are lossy and the transmitted photons decay exponentially over distance. To overcome this limitation, quantum repeaters can perform entanglement swapping to entangle distant parties.    

However, the intermediate quantum repeaters may be malicious. By establishing malicious entanglement, an intermediate node would jeopardize the confidentiality of the quantum information exchanged between two parties. This can be achieved by keeping the entangled pair secretly at the intermediate node instead of swapping them out, resulting in a man-in-the-middle (MitM) attack. Since qubits are closed systems until measured, neither party of the legitimate communication nodes can be certain that his/her part of the entangled pair is legitimately entangled with the other party. It is essential to emphasize that malicious entanglement poses a significant threat to the integrity of the exchanged quantum information, particularly when subjected to measurement in the wrong basis by a malicious repeater. Also, a malicious repeater can perform a denial-of-service (DoS) attack by directing the entanglement swapping to another node than the intended receiver. 

While Bell inequality tests and other entanglement-based tests play a crucial role in validating the integrity of entangled states within a quantum repeater network, their focus primarily revolves around confirming the quality of quantum connections. Hence, these tests do not directly incorporate mechanisms to ensure that the participants in communication are indeed the individuals they claim to be. Therefore, these tests fall short in comprehensively addressing the range of security challenges inherent to quantum communication networks. Notably, Bell inequality tests and entanglement-based tests do not provide a robust defense against insidious threats like impersonation or MitM attacks, which involve malicious parties posing as legitimate entities or intercepting the data exchange over the network. It is in this critical context that the significance of quantum identity authentication emerges, serving as a pivotal complementary mechanism. By directly verifying the identities of participating quantum nodes and detecting any unauthorized intervention, quantum identity authentication shores up the vulnerabilities left unaddressed by entanglement-based tests.
Thus, it is necessary to develop an effective authentication protocol that secures quantum communications against malicious entanglements.

\subsection{Related Works and Limitations}
\label{sec:relatedwork}
The authors in \cite{7} proposed a protocol for quantum identity authentication based on a shared classical secret key. However, some information may leak from the shared secret, and avoiding this would require authenticated classical channels. Also, in \cite{2}, a protocol is proposed for quantum identity authentication and key agreement by adopting random numbers to agree on the session key. The secret key is refreshed by the end of the protocol, which improves security but affects the efficiency of the protocol due to the associated overheads. Based on quantum key distribution, \cite{3} presents a quantum identity authentication protocol that can also be used to refresh the authentication key between the participants. Another authentication protocol is proposed in \cite{5} based on the BB84 version of quantum key distribution. In addition, \cite{4} proposes a quantum identity authentication protocol using a cluster state of five qubits. However, from a practical standpoint, it is not easy to prepare and maintain such cluster states. Moreover, a semi-quantum identity authentication protocol was proposed in \cite{1} using single-qubit measurement and XOR operations. Also, in \cite{46}, a protocol is introduced for quantum identity authentication by employing non-orthogonal states. Despite showcasing resilience against MitM attacks, it is important to note that the protocol's security is contingent upon the one-time utilization of authentication keys to maintain its effectiveness. Lastly, \cite{47} proposed a quantum identity authentication protocol centered on the utilization of quantum rotation properties and public key cryptography in a bit-oriented approach. While the protocol demonstrated commendable robustness against various attacks, the protocol of \cite{47} operates under the assumption of an ideal environment. Real-world imperfections within the quantum channel and detection mechanisms can directly influence the accuracy and security of the authentication process. Furthermore, potential losses within the communication channel may lead to authentication failures or create opportunities for malicious entities to exploit the noise for concealed attacks.

\vspace{2mm}

\noindent \textit{Limitations:} Most of the aforementioned quantum authentication protocols rely on quantum channels to transmit authentication qubits between the two communicating parties. Such channels are lossy and transmitted photons will decay over distance, which will reduce the success probability of the authentication process. 
Also, some protocols may leak information about the shared secret key, and hence, require a periodic refreshment of the shared secret keys.

\subsection{Contributions}

To overcome the existing limitations, the following contributions have been carried out: 
\begin{itemize}
    \item We propose a protocol that establishes authentication between communication parties in quantum networks. The proposed protocol relies on entanglement and reusable shared secret keys. 
    \item We perform a security analysis of the proposed protocol, which demonstrates that the detection probability of a malicious entanglement increases based on multiple authentication rounds from $50\%$ to reach $100\%$ by the fourth round of the authentication process. 
    \item We simulated the proposed protocol in a quantum network, which provided numerical results that support our security analysis. The simulations were carried out on a quantum network simulator, QuNetSim \cite{6}.   
\end{itemize}

The rest of this paper is organized as follows. Section \ref{sec:system} defines the problem of malicious entanglement in quantum networks and presents the proposed quantum identity authentication protocol and provides an illustration example. Section \ref{sec:results} presents a security analysis of the proposed protocol and discusses the simulation setup and results. Section \ref{sec:Conclusion} concludes the paper.   

\section{Problem Definition and Proposed Quantum Authentication Protocol}
\label{sec:system}

\subsection{Problem Definition}

Consider a quantum network represented as a graph $G=(V,E)$, where $V=\{v_i\}_{i=1}^N$ represents a set of $N$ nodes while $E=\{e_{i,j};v_i,v_j \in V\}$ represents the set of edges connecting the nodes.
Consider entangled pairs being already distributed among adjacent nodes. Let node $v_i$ want to send quantum data to a non-adjacent node $v_j$ and there is no direct quantum channel between $v_i$ and $v_j$. Hence, nodes $v_i$ and $v_j$ communicate via quantum teleportation. Being non-adjacent nodes, there is no entangled pair shared between $v_i$ and $v_j$. Consider an intermediate node $v_k$ that shares entangled pairs with $v_i$ and $v_j$ such that $e_{i,k} \in E$ and $e_{j,k} \in E$. Then, $v_k$ can perform entanglement swapping to directly entangle $v_i$ and $v_j$. Given the aforementioned model, this paper aims to provide a protocol that enables $v_i$ and $v_j$ to authenticate each other and ensure that $v_k$ did not perform a malicious entanglement. The proposed protocol works on any number of intermediate nodes and the adoption of a single intermediate node herein is meant only for illustration purposes.

\subsection{Proposed Quantum Authentication Protocol}
\label{sec:proposed}
The proposed authentication protocol assumes that the two end users share a secret key in advance, which can be done using a key management protocol such as quantum key distribution. Following the common terminology, let the end users be called Alice and Bob. Let Alice want to use teleportation to send quantum data to Bob. Through entanglement swapping via intermediate node(s), Alice and Bob can share an entanglement pair. Alice wants to authenticate Bob to ensure that she is sending the quantum information only to Bob and that the intermediate node(s) did not create a malicious entanglement. 

\subsubsection{Steps of the Proposed Protocol}

The proposed authentication protocol works as follows: 
\begin{enumerate}
    \item {Alice and Bob agree first on three parameters, namely, encoding index, base index, and transfer length. The encoding index is used to determine the initialization of an authentication qubit. The base index is used to determine the encoding and measurement bases of the authentication qubit. The transfer length is used to identify the number of data qubits that will be sent before an authentication qubit is sent. The encoding index and base index $\in \{0,1\}$ and must be the opposite of each other. For example, if the encoding index is $1$, then the base index must be set to $0$, and vice versa. For this reason, Alice and Bob do not need to agree on these two bits explicitly, they may agree on one of them and the other one will be set implicitly.}
    \item Alice and Bob use the transfer length and the shared secret key to calculate the number of data qubits that will be transferred before an authentication qubit is sent. The authentication process is repeated after every $R$ data qubits are transferred where $R$ is dynamic during a communication session between Alice and Bob and should be recalculated for each authentication round. For example, let $K$ be the shared secret key between Alice and Bob. Let Alice and Bob agree on a transfer length of $T$. Each of them will select the first $T$ bits from the secret key $K$ and calculate $R$ as their decimal equivalent. Then, Alice will send $R$ data qubits before waiting for an authentication qubit from Bob. Also, Bob will receive $R$ data qubits from Alice before sending an authentication qubit to Alice. Upon the reception of these $R$ qubits, the authentication process starts and after the authentication process is complete, Alice and Bob recalculate $R$ from the next $T$ bits of the secret key $K$. Hence, the number of data qubits sent before the authentication process changes after every authentication.
    \item {\label{step:repeat} Based on $R$, Alice keeps sending $R$ data qubits to Bob.}
    \item Bob keeps receiving $R$ data qubits from Alice.
    \item {Bob prepares an authentication qubit as follows:
    \begin{enumerate}
        \item {\label{substep:bob1} Bob sequentially chooses pairs of bits from the secret key, denoting the $i$-th two bits as the focal point. In the initial authentication round, this selection encompasses the first and second bits. Subsequently, in each successive round, the third and fourth bits take center stage, followed by the subsequent pairs in a similar manner. Notably, the first bit retains an index of $0$, while the second bit maintains an index of $1$ throughout each authentication round.}
        \item{\label{substep:bob2} Bob uses the encoding index and the base index to identify the initialization of the authentication qubit (encoding bit) and the encoding base (base bit) from the two bits selected at step (\ref{substep:bob1})}.
        \item{Bob creates an authentication qubit based on the identified values at step (\ref{substep:bob2}}). If the encoding bit is $0$, the authentication qubit is initialized to $\ket{0}$, otherwise, it is initialized to $\ket{1}$. If the base bit is $0$, the authentication qubit is encoded in the $Z$-bases, otherwise, it is encoded in the $X$-bases. Hence, the authentication qubit should be in one of the four states $\ket{0}$, $\ket{1}$, $\ket{+}$, or $\ket{-}$.
        \item Bob teleports the authentication qubit to Alice.
    \end{enumerate}}
    \item Alice receives the authentication qubit from Bob and decodes it as follows:
    \begin{enumerate}
        \item {\label{substep:alice1} Alice selects the $i$-th two bits from the secret key.}
        \item{\label{substep:alice2} Alice uses the encoding and the base indices to identify the measurement value and the measurement base from the two bits selected at step (\ref{substep:alice1}}).
        \item{Alice measures the authentication qubit based on the measurement base extracted at step (\ref{substep:alice2}), then, compares the result to the measurement value extracted at step (\ref{substep:alice2}). If they are not matched, then Bob's authentication has failed and Alice should terminate the session immediately.}
    \end{enumerate}
    \item{Alice and Bob calculate new $R$ and repeat the process from step (\ref{step:repeat}}).
    \item{If Bob wants to authenticate Alice, the same process is carried out in a reversed manner. But this means that Alice and Bob will sacrifice another entangled pair to do the reverse authentication.}
\end{enumerate}
\subsubsection{Illustrating Example}
\begin{figure}[!t]
    \centering
    \includegraphics[width=3.6in]{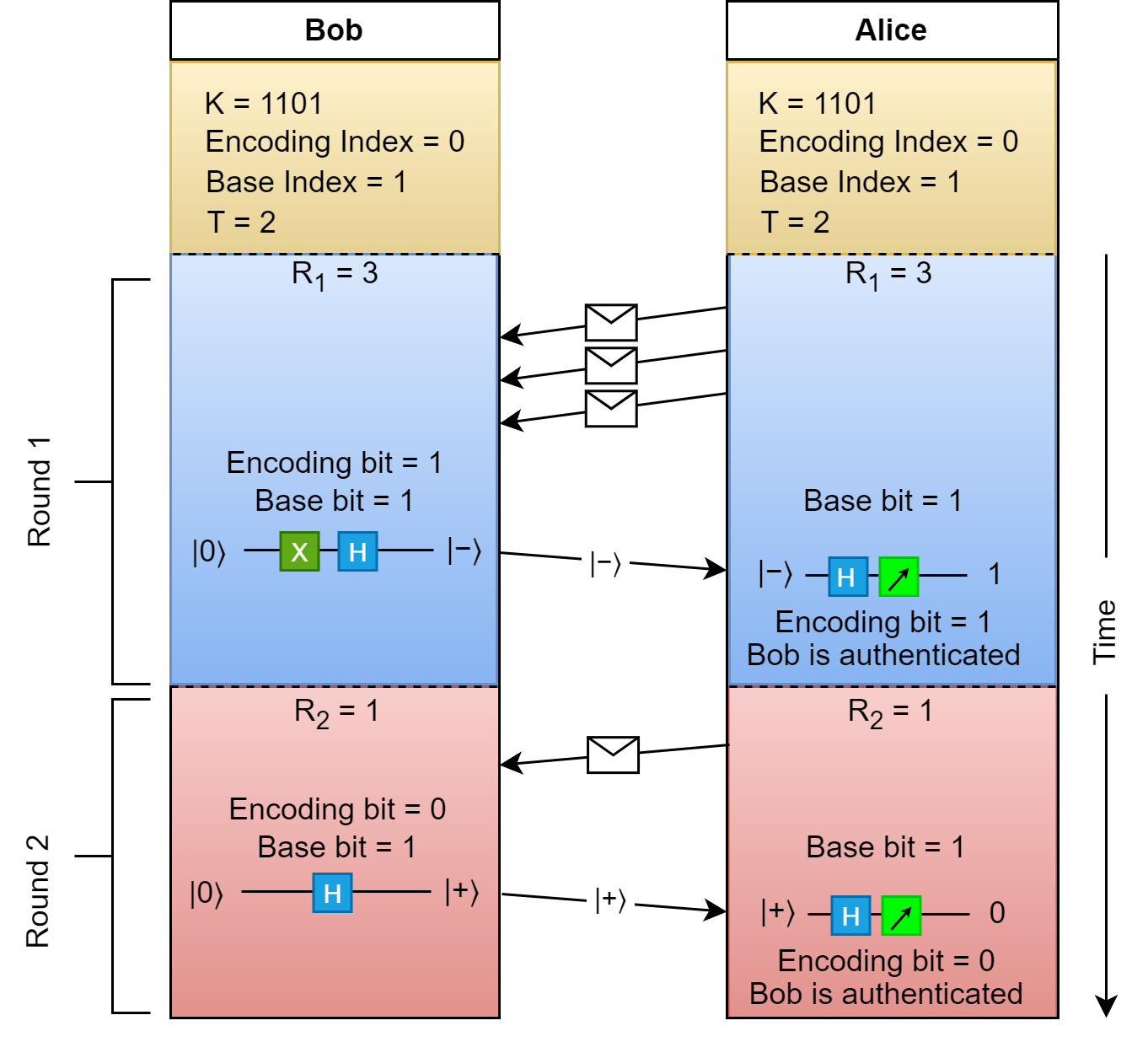}
    \caption{Illustration example of the proposed represented as timeline diagram.}
    \label{fig:auth_timeline}
\end{figure}

 To further explain the proposed protocol, the following example illustrates the first authentication round: Let Alice and Bob share secret key $K = 1101$ and agreed on an encoding index $0$, base index $1$, and transfer length $T = 2$. Fig. \ref{fig:auth_timeline} illustrates the steps of the proposed protocol as a timeline diagram.

\noindent \textbf{\textit{Step-1:}} Alice and Bob pick the first two bits from the key $11$ and convert them to decimal, which results in $3$. Alice and Bob calculate $R_1 = 3$. 

\noindent \textbf{\textit{Step-2:}} Based on $R_1$, Alice teleports $3$ data qubits to Bob before starting the authentication process. Also, Bob receives $3$ data qubits from Alice before starting the authentication.

\noindent \textbf{\textit{Step-3:}} 
The initial value and encoding base of the authentication qubit are extracted by selecting the first two bits from the key, where the encoding index is the first bit, and the base index is the second bit, and the bits are $11$. Bob then initializes the authentication qubit to $1$ in the X-bases using an X-gate followed by an H-gate, resulting in $\ket{-}$. Finally, Bob teleports the authentication qubit $\ket{-}$ to Alice.

\noindent \textbf{\textit{Step-4:}} 
Alice selects the first two bits from the key. The first bit is used as the encoding index, and the second bit is used as the base index. Alice measures the authentication qubit in the X-base and compares the result with the expected measurement value extracted from the key. If they match, authentication continues, and if not, Alice terminates the session immediately as this indicates the presence of an unauthorized party.

\noindent \textbf{\textit{Repeat:}} If Alice successfully authenticates Bob, they repeat the same steps until the session ends. In the second round, $R_2 = 1$. Hence, Alice sends $1$ qubit data to Bob before Bob starts to authenticate himself. When Bob authenticates himself, the authentication qubit will be  $\ket{+}$.   

\section{Security Analysis, Simulation Results, and Discussion}
\label{sec:results}
%\vspace{-2mm}
\subsection{Security Analysis}
\label{sec:analysis}
Assume Alice wants to send quantum data to Bob and an intermediate node (Eve) attempts to perform malicious entanglement. In this case, Alice and Bob are not directly entangled which means that Eve shares an entanglement pair with Bob and an entanglement pair with Alice. Alice and Bob share a secret key and Eve does not know the key. Suppose only Alice wants to authenticate Bob which is sufficient in some cases. Alice and Bob start the authentication process after a few transmitted qubits. Bob will prepare the authentication qubit and teleport it to Alice. In the malicious entanglement case, Eve will receive the qubit instead. Because Eve does not know the key, she will never know that this is an authentication qubit and will measure it. The arbitrary measurement performed by Eve on the authentication qubit will enforce the qubit to collapse in one of two possible states $\ket{0}$ or $\ket{1}$ and then Eve sends the qubit to Alice. The quantum information in the qubit prepared by Bob has been destroyed and Alice will detect the attack with a probability of $50\%$ after the first authentication round. This is because the authentication qubit has the chance to get collapsed to the correct value with a probability of $50\%$. After the second, third, and fourth authentication rounds, Eve will be detected with a probability of $75\%$, $87.5\%$, and $93.75\%$, respectively. Within $7$ authentication rounds, Eve will be detected with a probability of $99.2\%$. 

Also, Eve cannot impersonate Bob by constructing an authentication qubit because she does not know the key. So, she cannot know the initialization and encoding bases. Also, she does not know when Alice and Bob will perform the authentication because Alice and Bob do the authentication dynamically and secretly. 
The transfer length controls the amount of data transferred before each authentication. For example, a transfer length of $2$ can allow a maximum of $3$ qubits to be transferred before each authentication. A transfer length of $4$ can allow a maximum of $15$ qubits to be transferred before each authentication process. On one hand, if Alice and Bob transfer sensitive data, they may decrease the transfer length to do more authentication rounds and increase the probability to detect Eve before transmitting much data. On the other hand, Alice and Bob can increase the transfer length for higher throughput with fewer overheads if they are not transferring sensitive data.

Lastly, it is important to highlight that within their quantum communication setup, Alice and Bob do not rely on classical communication beyond the essential exchange of two classical bits needed for the recovery of teleported quantum states. This classical channel does not necessitate stringent security measures and remains unaffected by compromise.

\subsection{Simulation Results}
This paper used QuNetSim \cite{6} to simulate the proposed protocol and evaluate its performance. To test the security level of the proposed protocol, a malicious entanglement attack was simulated where Alice wants to communicate with Bob and there is a malicious node (Eve) performing a malicious entanglement attack between Alice and Bob. Alice shares an entangled pair with Eve, and Eve shares an entangled pair with Bob. It is supposed that Eve applies entanglement swapping to entangle Alice and Bob directly, but in this malicious scenario, Eve will not perform the entanglement swapping, which results in a MitM attack. The simulation scenario is tested for transfer lengths $T = \{1, 2, 3, 4, 5\}$, which means that the maximum number of data qubits transferred between two authentication rounds are $1, 3, 7, 15, 31$, respectively. For each transfer length, the simulation scenario is repeated $200$ times attempting to send $150$ data qubits from Alice to Bob. In each transfer, Eve receives a qubit from Alice, measures it in a random base, and then teleports it to Bob. 

\begin{figure}[!t]
\centerline{\includegraphics[width=3.35in]{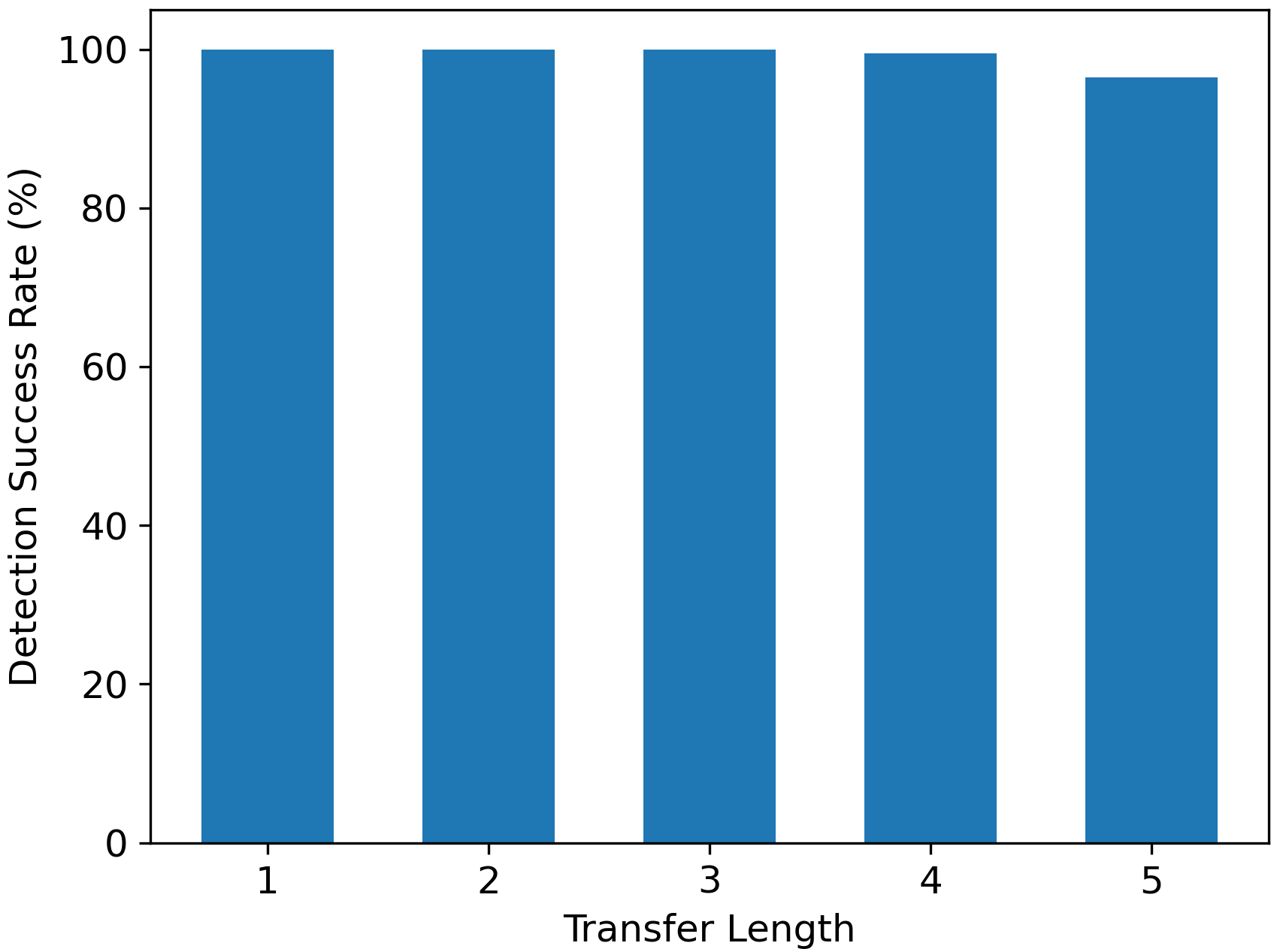}}
\caption{The success rate of the proposed protocol to detect a malicious entanglement attack, using transfer lengths of $1, 2, 3, 4,$ and $5$.}
\label{fig:success}
\end{figure}

\begin{figure}[!t]
\centerline{\includegraphics[width=3.35in]{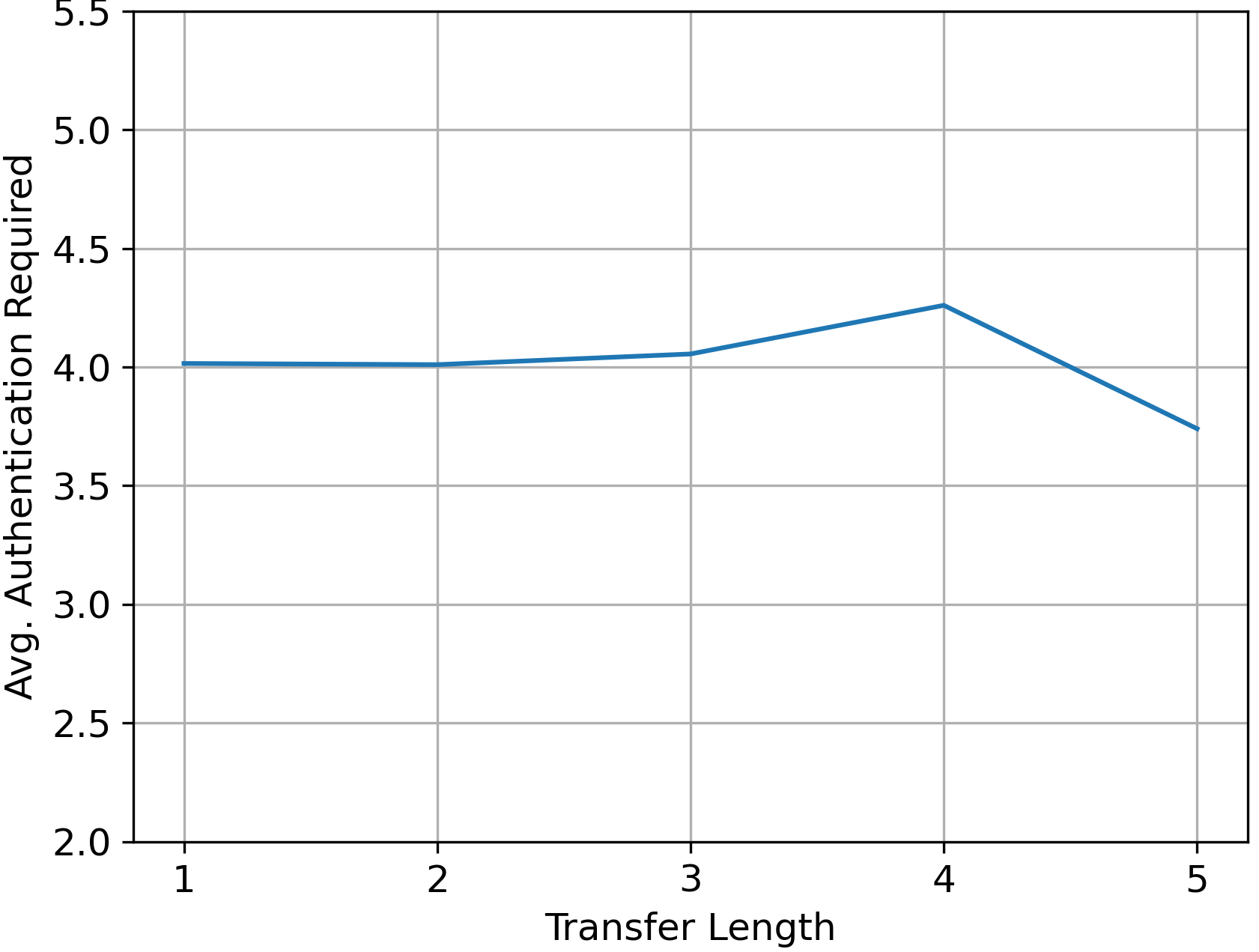}}
\caption{The average number of authentication rounds required for the proposed protocol to detect an attack, using transfer lengths of $1, 2, 3, 4,$ and $5$.}
\label{fig:auth}
\end{figure}

The success rate of the protocol is measured as shown in Fig. \ref{fig:success} where the proposed protocol is able to detect the attack with a probability of $100\%$ for the transfer lengths $1, 2,$ and $3$ while it degrades to $99.5\%$ and $96.5\%$ for the transfer lengths $4$ and $5$, respectively. This is because the number of authentication rounds performed is not enough compared with the amount of transferred data. The average number of authentication rounds required to detect an attack has been measured for each case as shown in Fig. \ref{fig:auth}. The simulation results demonstrate that the proposed protocol is able to detect an attack within $4$ authentication rounds on average. Although the security analysis presented in section \ref{sec:analysis} calculates that an eavesdropper would be detected within $7$ rounds with a probability of $99.2\%$, the simulation results showed that an eavesdropper can be detected with a probability of $100\%$ within just $4$ rounds. The security analysis focuses on calculating the worst-case scenario of detecting eavesdropping, while the simulation results calculate the average number of authentication rounds required to detect an attack within $200$ trials.

\begin{figure}[!t]
\centerline{\includegraphics[width=3.35in]{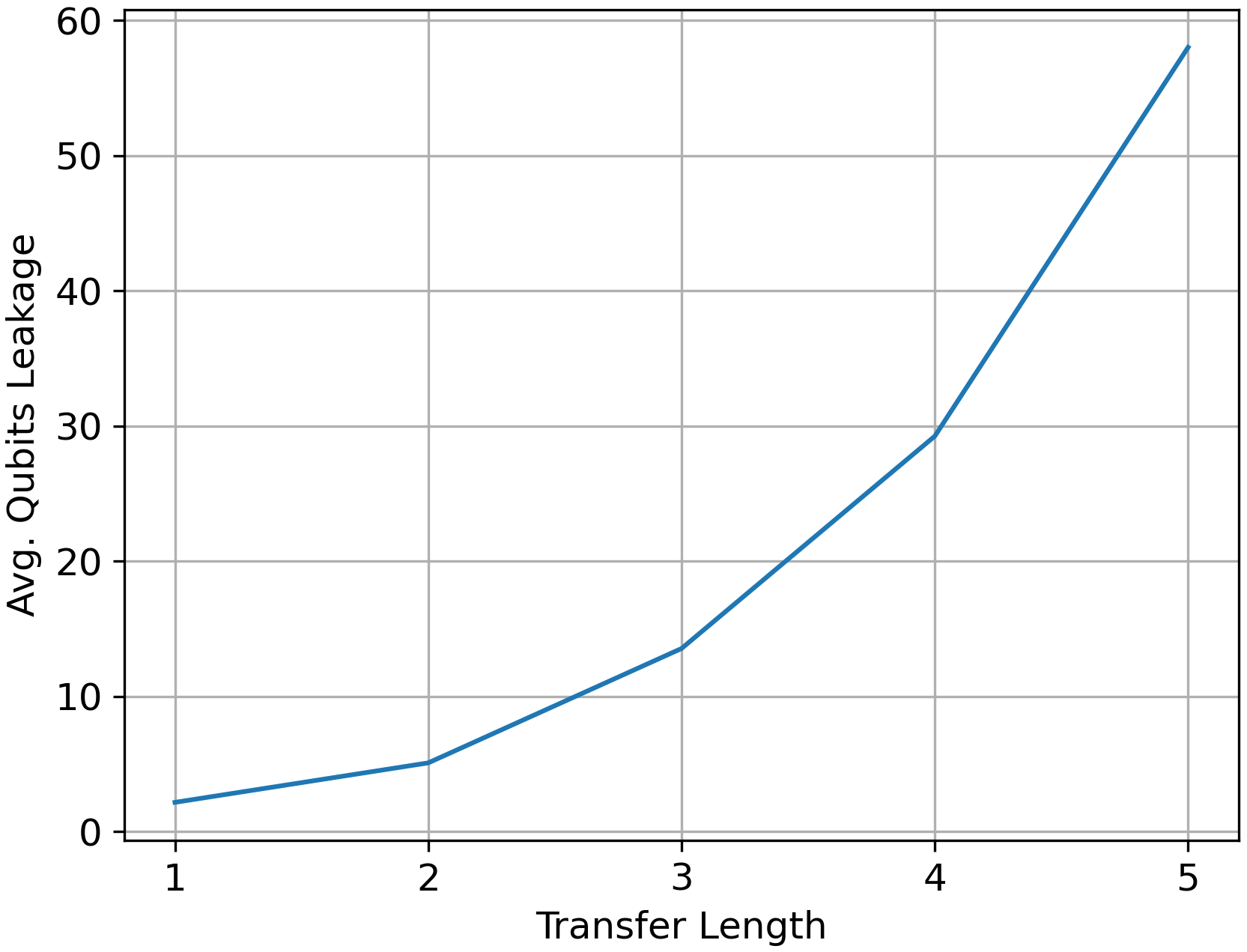}}
\caption{The average data leakage before detecting an attack and terminating the session, using transfer lengths of $1, 2, 3, 4,$ and $5$.}
\label{fig:leakage}
\end{figure}

The average data leakage has been measured for each case and shown in Fig. \ref{fig:leakage}. The data leakage is the number of qubits that are transmitted from Alice to Bob before detecting the presence of Eve. The results presented in Fig. \ref{fig:leakage} clearly demonstrate that the average data leakage increases as the transfer length increases. The underlying reason for this phenomenon is that as the transfer length increases, more qubits are transferred between two authentication rounds on average, providing Eve with more opportunities to gain information before being detected. Given the clear relationship between transfer length and data leakage demonstrated in Fig. \ref{fig:leakage}, it is important for Alice and Bob to carefully consider the confidentiality level of the data they are transferring when selecting an appropriate transfer length. If the data is highly sensitive, Alice and Bob should prioritize security over speed and choose a shorter transfer length. This will allow for more frequent authentication rounds, increasing the probability of detecting Eve before she can intercept and gather a significant amount of data. On the other hand, if the data is less sensitive, Alice and Bob may be able to afford a longer transfer length for faster communication. Ultimately, the decision of transfer length should be based on a careful balance between the confidentiality level of the data being transferred and the desired communication speed. By taking a thoughtful and strategic approach, Alice and Bob can ensure the highest level of security and protection for their data. The data leakage can lead to data integrity loss. Data integrity loss refers to the situation where the integrity or accuracy of data is compromised or altered, leading to inaccuracies, errors, or unauthorized modifications in the stored or transmitted data. It is essential to note that the data integrity loss significantly depends on the type of quantum data being exchanged. Without knowing the specific type of quantum data involved, providing accurate values for data integrity loss could be challenging. For instance, in scenarios where unknown quantum states are delivered and utilized without measurements, such as in distributed quantum computing, where qubits control another target qubit, the data integrity loss in this case is hard to quantify. This is because of the infinite number of quantum states a single qubit can represent if it exists in superposition. On the other hand, if the exchanged data involves classical data encoded as qubits, the integrity loss might be around $50\%$ of the data leakage. This occurs because the malicious repeater may measure the qubits in the wrong basis approximately $50\%$ of the time.

\begin{figure}[!t]
\centerline{\includegraphics[width=3.35in]{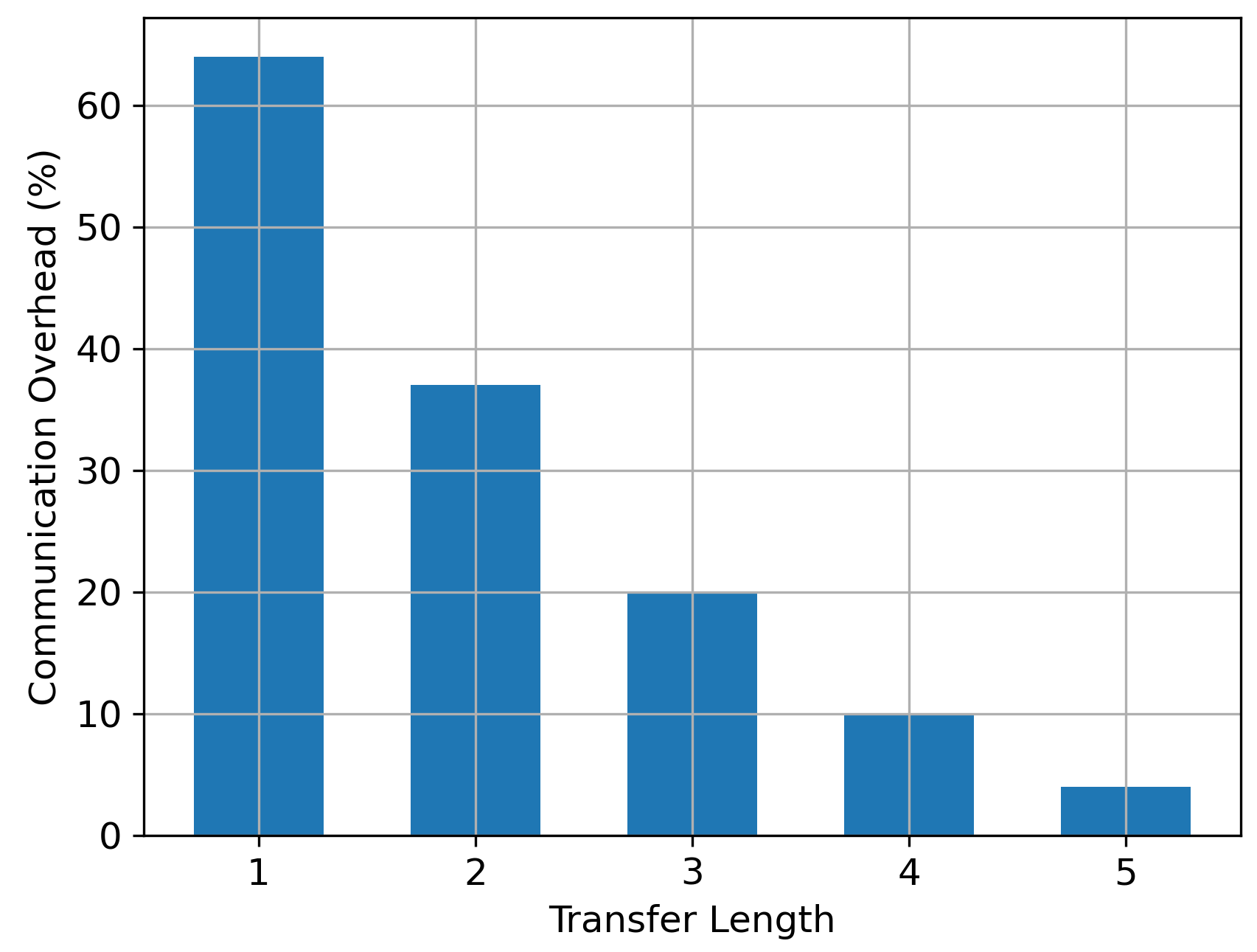}}
\caption{The average communication overhead of teleporting 100 qubits using the proposed protocol over transfer lengths of $1, 2, 3, 4,$ and $5$.}
\label{fig:overhead}
\end{figure}

To assess the practicality of using the proposed quantum authentication protocol, a simulation was conducted to measure the communication overhead of the authentication process. The communication overhead is defined as the ratio of the number of authentication qubits involved in the communication process to the number of data qubits. The simulation involved attempting to transmit 100 data qubits and the average of communication overhead was measured over 200 trials using transfer lengths of 1, 2, 3, 4, and 5. As shown in Fig. \ref{fig:overhead}, the communication overhead of the proposed protocol is $64\%$, $37\%$, $20\%$, $10\%$, and $4\%$ for transfer lengths 1, 2, 3, 4, and 5, respectively. These results demonstrate that while there is some overhead associated with using the proposed quantum authentication protocol, it is relatively small for reasonable transfer lengths. Therefore, the proposed protocol can be considered a viable and practical solution for secure communication between two parties.

The number of qubits that can be transmitted by a single use of the shared secret key depends on the transfer length. For example, if the shared secret key is $1024$ bits long and the transfer length is $2$, then the key can support up to $512$ authentication rounds. During each round, up to $3$ qubits of data can be sent, but on average only $1.5$ qubits are sent. Thus, the total number of qubits that can be sent by a single use of the $1024$-length key is $768$ qubits on average. If the transfer length is $3$, then the same key can support up to $341$ authentication rounds. During each round, up to $7$ qubits of data can be sent, but on average only $3.5$ qubits are sent. Thus, the total number of qubits that can be sent through a single use of the shared key is $1193$ qubits on average. 
Equation \eqref{eq1} can be used to calculate the average number $q$ of qubits that can be transmitted using a transfer length $T$ and a single use of a shared secret key with length $L$
\begin{equation}
    q = \lfloor \frac{2^T-1}{2} \times \frac{L}{T} \rfloor. 
    \label{eq1}
\end{equation}
However, if the shared secret key is used up, the two parties involved in the communication can start the process again from the first bit. Therefore, the number of qubits that can be sent is multiplied by the number of times the key is used. It is important to note that the shared secret key is highly reusable, as it does not reveal any information about the authentication process during transmission. This allows for a secure, covert authentication process that outside parties cannot detect. Because qubits are closed systems, it is impossible for anyone other than the communication parties to differentiate between the data qubit and the authentication qubit. Also, both the encoding bit and the base bit are fully secure, no one can determine the value encoded into the qubit or the encoding base used even if the authentication qubit is captured.


\section{Conclusion}
\label{sec:Conclusion}

This paper proposes a quantum identity authentication protocol to detect malicious entanglement over quantum networks.
To validate the effectiveness of the proposed protocol, comprehensive simulations have been conducted using the QuNetSim simulator. Specifically, an attack scenario has been designated to test the security level of the proposed protocol. The attack scenario is tested for transfer lengths of 1, 2, 3, 4, and 5 which mean that the maximum number of data qubits transferred between two authentication process are 1, 3, 7, 15, and 31 respectively. 
For each transfer length, the simulation scenario is repeated $200$ times attempting to send $150$ data qubits from Alice to Bob each time. In each transfer, Eve receives a qubit from Alice and measures it in a random base then transmits it to Bob. The results of each simulation have been recorded and the average has been calculated and illustrated in graphs to gain insights.
The simulation results show that the proposed algorithm can detect malicious entanglement with a probability of $100\%$ by $4$ authentication processes on average. The transfer length controls the amount of data transferred before each authentication. For example, a transfer length of $2$ can allow a maximum of $3$ qubits to be transferred before each authentication. A transfer length of $4$ can allow a maximum of $15$ qubits to be transferred before each authentication process. On one hand, if Alice and Bob transfer critical data, they may decrease the transfer length to do more authentication processes and increase the probability of detecting Eve before transmitting much data. On the other hand, Alice and Bob can increase the transfer length if they want faster communication and do not transfer critical data. 

In future work, the protocol will be enhanced to detect the denial of service attack that can be performed by malicious entanglement. Also, more attack scenarios will be studied and a security analysis will be provided for each attack scenario.

\bibliographystyle{IEEEtran}
\bibliography{references}
\end{document}